\documentclass[12pt]{article}
\newtheorem{theorem}{Theorem}
\def\re{\,{\rm Re}\,}
\def\im{\,{\rm Im}\,}
\newfont{\blackboard}{msbm10 at 12pt}
\def\lR{\mbox{\blackboard R}}
\def\lC{\mbox{\blackboard C}}

\begin{document}
\title{Quantum Formalism with State-Collapse and Superluminal Communication}
\author{George Svetlichny\thanks{ Departamento de Matem\'atica,
Pontif\'\i cia Universidade Cat\'olica, Rio de Janeiro, Brazil, e-mail:
svetlich@mat.puc-rio.br}}
\date{\today}
\maketitle
\begin{abstract}
Given the collapse hypothesis (CH) of quantum measurement, EPR-type
correlations along with the hypothesis of the impossibility of
superluminal communication (ISC) have the effect of globalizing gross
features of the quantum formalism making them universally true. In
particular, these hypotheses imply that state transformations of density
matrices must be linear and that evolution which preserves purity of
states must also be linear. A gedanken experiment shows that lorentz
covariance along with the second law of thermodynamics imply a
non-entropic version of ISC. Partial results using quantum logic suggest,
given ISC and a version of CH, a connection between lorentz
covariance and the covering law. These results show that standard
quantum mechanics is structurally unstable, and suggest that viable
relativistic alternatives must question CH. One may also speculate that
some features of the hilbert-space model of quantum mechanics have their
origin in space-time structure.
\end{abstract}

\section{Introduction}\label{sec:intro}
Inspired by the paper of Einstein, Podolsky, and
Rosen \cite{einstein:can}, and more recently by the controversy
surrounding the Bell inequalities \cite{bell:ontheeinstein,
redhead:incompleteness}, many researchers have tried to interpret
long-range quantum correlations as resulting from some sort of
action-at-a-distance. A natural question then arises as to whether such
correlations can be used for superluminal communication. Several
authors \cite{eberhard:two, eberhard:three,ghirardietal:onsome,
ghirardietal:nohave2} argued that no such communication is possible
since the statistical behavior of any detector placed on one arm of an
EPR apparatus is completely independent of what is done on the other
arm. Nick Herbert \cite{herbert:nohave} argued that if one can construct
a ``photon duplicator" which reproduces exactly an incoming photon
state, then superluminal communication is possible. This argument
quickly provoked several rebuttals \cite{dieks:communication,
milonietal:photons,wootersetal:asingle} to the effect that no linear
state transformer can clone an arbitrary photon state. It seems
therefore that ordinary hilbert-space quantum mechanics precludes the
use of EPR-type correlations for superluminal communication. We address
in this section the reciprocal question: under the hypothesis of the
impossibility of superluminal communication, what can one deduce about
the behavior of detectors and state transformers? It is a common view
that superluminal communication should not be possible as it raises
serious questions of relativistic covariance and causality. These will
be further considered in Section~\ref{sec:thesecond}.

To be able to attack the question at hand we shall initially work with
the hypothesis that there are possibly some physical processes that may
not conform to the usual quantum mechanical description but that these
are specific to very particular situations, whereas for the vast
majority of other processes (including all those experimentally studied
up to now), any deviation from normal quantum mechanical predictions is
below present experimental precision. Thus one may posit that there may
be a state transformer, such as the hypothetical photon cloner mentioned
above, that acts in a non-linear fashion, and that such a transformer
may take part in an experimental arrangement in which normal quantum
mechanical description is adequate for processes not involving it.
Explicitly the hypothesis is then that, in any given inertial frame, up
to the use of an unconventional device, the usual quantum mechanical
reasoning can be used, including the projection rule. Up to such a
moment, ordinary quantum mechanics determines what the physical state
is. At the point of using the unconventional device we of course must
posit what would happen (a photon would be cloned in the above cited
example). We shall call allowing the situations described above the {\em
neighborhood hypothesis\/} (NH) since what we are describing, informally
speaking, is a situation that would neighbor standard quantum mechanics
in the set of all possible physical theories.

We shall show in sections \ref{sec:theprojection} and \ref{sec:quantum}
that under NH, certain types of deviation, specifically non-linearities
and lack of true randomness of outcomes,  allow for superluminal
signals. These theories are thus to be ruled out if we assume the
impossibility of superluminal communication (ISC). This makes ordinary
quantum mechanics a structurally unstable theory.  This is important as
many proponents of modifications to ordinary quantum mechanics are in
fact implicitly assuming NH and so face a real risk of coming into
conflict with special relativity.

It must be emphasized that NH is essentially an assumption about
formalism and not about interpretation.  It is an alteration of the
formalism that is generally part of what is known as the Copenhagen
interpretation, but we make no interpretational hypotheses. Though state
collapse is used, we make no assumption as to its ontological nature,
only that it is a legitimate calculating device for joint probabilities
of events. In the end what we are saying is that joint probabilities
cannot be calculated by certain rules if ISC is to be maintained. This
makes our results basically interpretation independent.

It should also be noted that part of our understanding about the
standard formalism is that it is capable of giving account of a
relativistically covariant theory. This is not straightforwardly obvious
given the instantaneous nature of wave function collapse
 \cite{aharonovetal:nohave, aharonovetal:istheusualii}, but this does not
preclude lorentz covariance of observable quantities. What the standard
formalism lacks is thus {\em manifest\/} covariance while being able to
provide for covariance of measurable magnitudes. It is precisely this
fact that makes the theory structurally unstable, for a perturbation in
the formalism is likely to make the manifest non-covariance capable of
producing real effects, such as superluminal communication.

In section \ref{sec:thesecond} we relate ISC and the second law of
thermodynamics showing that under certain hypotheses superluminal
communication can be used to foil the second law. In section
\ref{sec:thecovering} we argue that ISC, along with the projection
postulate, can be used as supporting evidence for assuming certain
axioms in the foundations of quantum mechanics thus suggesting that
quantum mechanics owes some of its aspects to space-time structure. In
the last section we draw some brief conclusions from these
considerations.

\section{The projection postulate and superluminal communication.}
\label{sec:theprojection}

        We shall assume the existence of an EPR-type apparatus by which
a physical system decomposes into two parts which then separate in such
a way that future measurements on each part separately can be performed
at space-like separation. To avoid the complications of Bose or Fermi
symmetrization, we shall assume the two parts are not identical,
nevertheless we posit that the internal degrees of freedom of each part
are described by finite-dimensional hilbert spaces of the same dimension
which is at least three. In what follows we assume that the wave
function factors into a product of the spatial part and the internal
part. We focus only on the internal factor. Let \(e_1,\dots,e_N\) and
\(f_1,\dots,f_N\) be orthonormal bases for the hilbert spaces of the
internal degrees of freedom. We shall then assume that one can prepare
the composite system in the state \(\Psi=(1/ \sqrt N)(f_1\otimes
e_1+f_2\otimes e_2+\cdots+f_N\otimes e_N)\). We also assume that for one
arm of the apparatus (hereafter referred to as arm A), given any basis
\(h_1,\dots,h_N\) for the corresponding hilbert space, we can measure a
non-degenerate observable with this eigenbasis. Swift and
Wright \cite{swiftetal:generalized} have argued that for certain spin
systems one can physically prepare a state corresponding to any ray in
the corresponding hilbert space, and construct an actual apparatus
corresponding to any self-adjoint operator. We shall thus generally
assume that there are apparatus corresponding to any self-adjoint
operator that we consider. We chose a reference frame in which the
observation on arm A is temporally prior to that on arm B and, under the
Neighborhood Hypothesis, and the assumption that superluminal
communication is impossible, determine what must hold at arm B.

         We first note that \(e_i = \sum u_{ij}h_j\) for some unitary
matrix \(u\). This means that \(\Psi =(1/ \sqrt N) \sum g_j\otimes h_j\)
where \(g_j = \sum u_{ij}f_i\) defines another orthonormal basis, and
any basis can be so constructed. Under the usual projection postulate,
if we now perform observations corresponding to a non-degenerate
self-adjoint operator  with eigenbasis \(h_1,\dots,h_N\) on arm A, then
subsequently the state collapses to a mixture in equal proportions of
product states \(g_j\otimes h_j\), and consequently the part of the
system that is in arm B can be construed as being, in equal proportions,
in the well defined quantum states \(g_j\). Let us now locate a detector
on arm B of the apparatus and suppose that the detection rate for a pure
state \(\phi\)  is \(D(\phi )\). The detection rate on arm B is then
\((1/N)\sum D(g_j)\). For superluminal communication to be impossible
this number must be independent of the orthonormal basis
\(g_1,\dots,g_N\) since otherwise the person stationed on arm A could,
by changing his observable, induce a change in the detection rate on arm
B at a space-like separation. This means by Gleason's
theorem \cite{gleason:nohave}, that there is a positive operator \(R\)
such that \(D(\phi ) = (\phi,R\phi )\). Since an observable can be
construed as the simultaneous action of a number of mutually exclusive
compatible detectors (one for each ``pointer position"), we also
conclude that on arm B, the mean value of any observable in state
\(\phi\) must be given by the usual quantum mechanical formula \((\phi ,
A \phi )\) for some self-adjoint operator \(A\). As far as the
expectation values of observables are concerned, arm B must then also
follow the usual rules. Let us now consider on arm B a state transformer
that transforms a state \(\phi\) into another, possibly mixed state. Let
us indicate this action on the corresponding density matrices as
\(\rho_\phi \mapsto T\rho_\phi\), where \(\rho_\phi  = (\phi
,\cdot)\phi\) and \(T\) is {\it a-priori\/} an arbitrary map. If, after
we subject the incoming state to transformation \(T\), we then perform
an observation corresponding to a conventional quantum observable
represented by the operator \(A\), the resultant expected value is
\(Tr(AT\rho_\phi )\) and by our previous result this must be of the form
\((\phi ,\tau(A)\phi )\) for some self-adjoint operator \(\tau(A)\).
Choose now an orthonormal basis \(k_1,\dots,k_P\) for the space in which
the transformed states lie (which may be different from the space in
which \(\phi\) lies, as would be the case for the putative ``photon
cloner"). Setting \(A = (1/2)((k_p,\cdot)k_q+(k_q,\cdot)k_p)\) and then
\(A = (1/2i)((k_p,\cdot)k_q-(k_q,\cdot)k_p)\) one deduces that \(\re
(k_p, T\rho_\phi  k_q)  = (\phi , M_{pq} \phi )\) and \(\im (k_p,
T\rho_\phi  k_q) = (\phi ,N_{pq} \phi )\) for some self-adjoint
operators \(M_{pq}\) and \(N_{pq}\) with \(M_{pq} = M_{qp}\) and
\(N_{pq} = -N_{qp}\). If we now set \(L_{pq} = M_{pq} + iN_{pq}\) we
have \((k_p, T\rho_\phi  k_q) = (\phi ,L_{pq} \phi )\) and finally
\[T\rho_\phi  = \sum (\phi , L_{pq} \phi )(k_q,\cdot)k_p = \sum
Tr(L_{pq}\rho_\phi )(k_q,\cdot)k_p\] but this is a {\em linear\/}
function of \(\rho_\phi\). We thus conclude that at arm B any density
matrix state transformer must be linear. Finally, suppose the state
transformer does not turn pure states into mixed states, then we can
write the transformation as \(\phi  \mapsto S\phi\). Now by the previous
result one has that \((\phi ,\cdot)\phi  \mapsto (S\phi ,\cdot)S\phi\)
must be a restriction of a linear map and given by the formula above. We
have been able to prove a theorem (see the appendix) which states that
in this case, if the range of \(S\) is not confined to one ray, \((S\phi
,\cdot)S\phi  = (C\phi ,\cdot)C\phi\) where \(C\) is either a linear or
an anti-linear t ransformation. Non-degenerate state transformers on arm
B that do not create mixed states from pure ones must therefore be
representable by either linear or anti-linear mappings. Summing up:

\begin{quote} {\em Under the Neighborhood Hypothesis and normal quantum
mechanical processes in relation to one arm of an EPR-type apparatus,
and assuming the impossibility of superluminal communication, then at
the distant other arm 1) the mean value of observables must be given by
the usual quantum mechanical expectation value formula 2) density-matrix
state transformers must be linear and 3) vector state transformers whose
range contains more than one ray must be either linear or anti-linear.}
\end{quote}

        The wayward transformers are those of the form \(S\phi =
\theta(\phi) (\phi,D\phi)^{1/2}\psi\) for some positive operator \(D\),
vector \(\psi\), and unimodular function \(\theta\). Note that
\((S\phi,AS\phi)\) is just \((\phi,D\phi)(\psi,A\psi)\) and so no
measurement on \(S\phi\) provides any more information about \(\phi\)
than that already given by the matrix elements \((\phi,D\phi)\). Hence
the transformer is no more than the observable \(D\) in disguise, no
real information about \(\phi\) being passed on to the transformed state
\(\psi\). This degenerate case is thus of little further interest.

The analysis of this section doesn't go through for quantum systems
described by a two-dimensional Hilbert space as Gleason's theorem does
not then hold. Nevertheless, as Nick Herbert's  \cite{herbert:nohave}
example shows, certain non-linearities can still lead to superluminal
communication. We have not determined what the exact restrictions on
non-linearities are in the two-dimensional case as these are specialized
physical situations. Although logically tenable, the idea that physics
changes radically once a physical system described by a two dimensional
Hilbert space gets isolated from the environment, is to allow what must
be deemed a rather bizarre situation.  From the still informal viewpoint
of a neighborhood of standard quantum theory, such theories would not
constitute a generic situation and are thus left out of this  first
analysis.

        What this all suggests is that some of the gross features of the
quantum mechanical formalism are intimately connected to spatio-temporal
relations. In particular, if part of the world follows standard rules,
then the rest must follow suit if superluminal communication is to be
ruled out. There cannot be any ``small deviations". To what extent the
impossibility of superluminal communication, {\em per se\/}, limits any
possible physical theory is not clear. One is aware of the great
difficulties in trying to justify the quantum formalism on clear {\em
a-priori\/} physical grounds. The existing axiomatic schemes, of which
the two most developed are due to Ludwig \cite{ludwig:nohave} and
Piron \cite{piron:foundations}, all suffer from the defect that the
crucial axioms are by no means compelling nor even clear as to their
true physical content. There is now the intriguing possibility that some
spatio-temporal hypothesis such as the impossibility of superluminal
communication could result in hilbert-space quantum mechanics in a more
natural way. We begin to address this question in
Section~\ref{sec:thecovering} of this paper.

        Various researchers have expressed the hope that some paradoxes
of quantum mechanics (especially those connected with ``Schr\"odinger's
Cat", that is, measurement theory) would eventually be resolved by
non-linearities (which presumably become more important as the number of
particles increases) in the time evolution \cite{pearle:reduction}. We
see now that any such non-linearity carries with it the real possibility
of superluminal communication and its concomitant problems. Steven
Weinberg \cite{weinberg:precision, weinberg:testing} has proposed a
non-linear quantum theory which in principle could be tested
experimentally by effects such as spectral line broadening. Any such
effect could be immediately used to build a superluminal communication
device. That this is possible for the Weinberg theory has already been
shown by Gisin \cite{gisin:weinberg} who has independently come to
conclusions similar to the ones presented in this paper. That there is a
connection between quantum evolution and superluminal communication has
been pointed out in various articles by Gisin \cite{gisin:quantum,
gisin:reply, gisin:stochastic,gisin:weinberg} and by
Pearle \cite{pearle:comment, pearle:stochastic}. Our result shows the
generality of such a connection, it essentially is a consequence of the
projection postulate and the structure of hilbert space, concretely
Gleason's theorem.

Deeming superluminal communication undesirable, one can try to avoid it,
while maintaining non-linearity, by modifying the projection postulate.
While logically plausible, there are  at present no known examples of a
relativistic non-linear quantum mechanics. We discuss
elsewhere \cite{svetlichny:quantum} the possibility of such a theory. In
recent years there has been a growing interest in investigating
non-linear evolution in spite of the reasons, such as those discussed in
this paper and elsewhere \cite{svetlichny:nonlinear}, that have been
brought forth against it. The feeling seems to be that an appropriate
reformulation of the measurement process would eliminate the
difficulties. Investigations by G.~A.~Goldin, H.-D.~Doebner, and
P.~Nattermann \cite{goldin:nohave,
doebneretal:nohave1,doebneretal:nohave2} show that by a reasonable
restriction on the set of allowed measurements, certain non-linear
Schr\"odinger equations are then observationally equivalent, via a
non-linear ``gauge transformation" to the free linear equation. This
shows that non-linearity {\em per se\/} may not lead to superluminal
signals under a reasonably modified measurement postulate. As the
equation studied by these authors are non-relativistic, we are still far
from understanding the true relation of linearity to relativity.

\section{Quantum indeterminism and superluminal communication.}
\label{sec:quantum}

        The so called ``quantum indeterminism" also seems to be
connected with spatio-temporal relations. To see this, still under the
Neighborhood Hypothesis, consider the already familiar EPR situation for
the singlet two-photon states: there is a photon in each arm of the
apparatus, and the combined state is the singlet: \((1/\sqrt 2)(H\otimes
V+V\otimes H)\) where \(V\) indicates vertical linear polarization and
\(H\) the orthogonal horizontal one. Consider a linear polarizer. The
usual assumption is that the passage of a non-polarized photon  (such as
one of the photons of the singlet state) through this polarizer is
absolutely random. If we write down a sequence of \(0\)'s and \(1\)'s
where \(0\) means the photos is absorbed, and \(1\) passed through, then
this sequence is thought to be at least a von Mises random
sequence \cite{schnorr:nohave} in that no computable function that tries
to predict the next outcome on the basis of outcomes already seen does
better than chance. Suppose now that besides such a ``randomizing"
polarizer one also has a non-randomizing one for which some computable
predictor does better than chance. Let one person then stay on arm A of
the EPR apparatus armed with the two kinds of polarizers, which he may
place in a horizontal orientation. Let his colleague stay at arm B with
a polarizer placed in the vertical position, and let him use the
computable predictor to guess whether the photon passes or not. Due to
the strict mirror correlations in the singlet state, the sequence seen
at arm B is the same one that is seen at arm A. Suppose that at arm A
the randomizing polarizer is placed, then at arm B the predictions are
seen to be no better than chance. Now one places the non-randomizing
polarizer at arm A and shortly thereafter one sees at arm B better than
chance predictions. If the arms are sufficiently distant this
constitutes superluminal communication. Thus if quantum indeterminism
exists in part of the world it must be universal if superluminal
communication is to be ruled out. Once again, under the assumption of
the impossibility of superluminal communication, distant quantum
correlations have the effect of universalizing gross features of quantum
mechanics. As before, it is not at all clear at this stage if the
impossibility of superluminal communication, {\em per se\/}, imposes von
Mises randomness on quantum events.

\section{The second law of thermodynamics and superluminal
communication}\label{sec:thesecond}

        In Sections \ref{sec:theprojection} and \ref{sec:quantum} we
showed that assuming the impossibility of superluminal communication,
and given the projection postulate, one can deduce that certain gross
features of the quantum formalism must be universal and not admit any
deviation. In this section we explore the relation of the hypothesis to
thermodynamics by showing that given lorentz covariance, the second law
of thermodynamics precludes non-entropic superluminal communication.

        The second law and the hypothesis of the impossibility of
superluminal communication have many superficial similarities. The most
important of these is that both are statements about general physical
objects of arbitrary size and complexity, stating that no such object
can perform a certain task. In the case of the second law the task would
be (among many equivalent formulations) to extract useful energy from
heat, in the second case that of transmitting a message across a
space-like interval. Superluminal communication {\em per se \/} is
innocuous and only becomes problematic if its underlying physics is
required to be lorentz covariant. In this case, it is well known that
one can then send a message from a point on a time-like world-line to
another point on the same world-line in the causal past of the first one
(communication to one's past).  It would then be possible for me to
receive a message signed by me and dated tomorrow that reads: ``Under no
circumstances send this message!" Believing in free will, I take a firm
resolve to obey the request. Am I then coerced in spite of my resolve to
send it? If so, what happened to free will? If not, how did I happen to
receive it? Superluminal communication in a lorentz covariant world is
fraught with such causality paradoxes. In what follows we shall
hypothetically assume such retrograde messaging in contexts where the
paradoxical nature is either minimized or absent, tacitly supposing that
its more bizarre manifestations get somehow resolved or avoided through
mechanisms or circumstance yet unspecified. Presumably a consistent
lorentz-covariant mathematical theory involving superluminal
communication might resolve the paradoxes by imposing adequate boundary
conditions on the solutions of the dynamic equations.

        The second law is also about entropy, saying that it does not
decrease in closed systems. Entropy is intimately tied to information.
Communication is the transfer of information. It is then not surprising
that if information can flow from the future to the past, reversing its
normal direction, then entropy in a closed system where such messaging
takes place can be made to decrease, also reversing its normal
direction.

        In contemplating the thermodynamics of a system involving
information transfer, it is necessary to take into account the entropy
created in the process. Now a functioning communication channel should
not in principle create entropy. To be clear, we are here considering
the channel mechanism itself and not the processes involved in getting
the message into and out of the channel. All such known channels involve
the transfer of energy or matter (think of television or the postal
service) and actually create entropy, but this entropy is incidental to
the actual information transferred and comes about through
inefficiencies that in principle can be eliminated or  minimized below
any preassigned level. One can communicate by a conservative and purely
mechanical device. All that is needed is that the observed motion,
invoked by the sender, have a meaning to the receiver. There are no
known channels of superluminal communication. The hypothesized schemes
or involve tachyons whose existence has not been established, or
EPR-type correlations which cannot be used for communication unless, as
was shown in Section~\ref{sec:theprojection}, the hilbert-space quantum
formalism breaks down, for which likewise there is no evidence. But even
in such hypothesized channels there is no evident reason for energy
degradation to be necessary. We shall therefore assume initially that
our putative superluminal communication process does not create entropy,
or at least that such entropy can, for a given message, be reduced as
much as necessary. There {\em is \/} though some unavoidable entropy
creation involved in {\em observing\/}  a system (to get the necessary
information that will become the content of the message) as has been
pointed out by Brillouin \cite{brillouin:maxwell, brillouin:physical}. As
the neglect of this type of entropy would lead to second-law violation,
via a Maxwell's demon type argument, even without positing superluminal
communication, these entropy sources must be considered.

        We shall begin our discussion then with the assumption of a {\em
non-entropic\/} superluminal communication process whose underlying
physics is lorentz covariant. We then construct a communicator that
sends messages to the past. This generally is achieved by signaling
between different inertial frames, but this can be done by conventional
means and so we can continue to posit that no entropy is created.

        There are many (unworkable) proposals to foil the second law
using thermodynamic fluctuations and one-way mechanical devices
(ratchets, valves, etc.). The idea is that the device would be activated
by a fortuitous but inevitable fluctuation, and the one-way nature of
the device would prevent a reverse motion thereby storing energy (in a
spring say) which is then extracted and used. One need only then wait
for the next fluctuation to continue the process of extracting useful
energy from heat. All such schemes fail. They {\em seem \/} plausible
since one generally overlooks details of exactly what happens when the
fluctuation activates the device. Surprisingly enough many, if not all,
such proposals can be made to work if one can anticipate events, and
retrograde communication makes this possible. Let us contemplate then
one such proposal. Consider an enclosed rectangular volume with a rigid
partition in the middle dividing the volume into two chambers (denoted
by A and B) filled with a gas at the same pressure and temperature. Let
us put a small one-way valve on the partition, the mobile element of
which is held in place by a spring. If the local pressure on side A of
the valve exceeds that on side B by some threshold difference then the
force of the spring is overcome and the valve opens. Pressure excess on
the other side has no effect due to the one-way nature of the mechanical
construction. The (fallacious) argument is that due to pressure
fluctuations the valve would open from time to time and  allow gas
molecules to pass from the supposedly (local) higher pressure side to
the lower pressure side transferring molecules from chamber A to B and
building up a true pressure (and temperature) difference violating the
second law. The argument falls down because the molecules that push the
valve open have already recoiled and are not the ones that actually pass
through it. These, passing from both sides, have such a distribution of
velocities that no pressure difference is built up. If however it were
known beforehand {\em when \/} the valve would be pushed open, then one
can simply temporarily remove the valve, exposing thus an orifice on the
partition, and then the exchange of molecules across the orifice would
be such that those passing from A to B would have a larger mean velocity
than those passing in the other direction. Such information can be had
if retrograde communication is possible and the second law can then be
violated. This is essentially a variant on the Maxwell's demon scheme.
The usual Maxwell's demon cannot violate the second law, since the
process of observing the velocity of molecules, if done by actual
photons, creates more entropy than the decrease achieved in operating
the orifice closure to selectively let molecules pass from one side to
the other \cite{brillouin:maxwell}. The observing photon is necessarily
of higher energy than the mean kinetic energy of the molecules and this
thwarts the attempt. Our demon (observer) does not observe the
molecules, all he need become aware of is a normal consequence of a
fluctuation, such as a valve opening, and this can be done with low
energy photons since the dimensions of the valve are much larger than
those of the molecules and it's motion is much slower.

        This scenario does have its paradoxical aspects (inevitable
consequence of the conjunction of lorentz covariance and superluminal
communication). If the observer has free will and receives the correct
information that at some instant the valve will be pushed open, then by
removing  the valve during a small time interval he destroys the chance
of observing the valve being activated. He can still send the message to
his past to maintain consistency with the fact of having received it
(not doing so creates an even bigger paradox) but then the message has
no real basis in observed events. This is self-consistent but hardly
satisfying. To maintain some semblance of the ordinary order of things
one can assume that the valve itself has an aperture that can be opened
and closed and that it is small relative to the size of the valve so
that even with the aperture open, pressure fluctuations still activate
the valve. The observer now only send a message if he sees the valve
activated (having already also opened the aperture by request of the
message received). This still works as fluctuations have a certain size
which gives a greater probability to transferring, via the aperture,
higher velocity molecules from A to B and lower velocity ones in the
opposite direction than to doing the opposite. Having reached this stage
we can now actually eliminate the  Brillouin ``observer." One can
automate the whole communication and valve operating process. The
mechanism that receives the retrograde message can be coupled directly
to the mechanism that opens the orifice. The message content itself can
be reduced to a single bit, whose meaning would be ``exactly \(T\)
seconds from now open the orifice". One would need to keep an accurate
clock and operate a device based on it, but this again in principle need
not create entropy. The whole gadgetry surrounding the retrograde
communicator can be taken to be a conservative system. The act of
observing the valve can be eliminated by mechanically coupling the valve
mechanism itself to the communication system. At this point the
paradoxical nature of retrograde communication is practically absent.
Our Rube-Goldberg-like gedanken experiment should of course be taken as
a sort of visualizable enactment of a physical process that would be
governed by a precise mathematical theory (using both advanced and
retarded potentials, say), if indeed it were desirable to construct such
a theory.

        The assumption that the superluminal communication channel does
not create entropy is crucial to the above argument. One may think that
this hypothesis is not essential, arguing that what one need do is
simply operate the scheme in those circumstances where the entropy
decrease achieved (by anticipating an adequate fluctuation, say) is
greater than the entropy gain in using the channel. Although with
specific assumptions about entropy created by the  channel one can carry
this through, there doesn't seem to be any {\em general \/} argument, at
least we haven't been able to find one. We therefore tentatively
conclude that one must necessarily impose restrictive assumptions on
entropy creation in the channel in order to violate the second law.

        Let us now summarize the hypotheses that we are relating: (LC),
lorentz covariance; (SL), the second law of thermodynamics; (ISC), the
impossibility of superluminal communication; (INESC), the impossibility
of non-entropic superluminal communication, and (UQFCH), the
universality of the quantum formalism with the collapse hypothesis.

        We have finished showing: \((LC \> \> {\rm and} \> \> \sim INESC
\> \Rightarrow \> \> \sim SL)\) which we prefer to state as:  \((LC \>
\> {\rm and}\> \>  SL \> \Rightarrow \>  INESC)\). In
Section~\ref{sec:theprojection} we have proved: \((ISC \> \Rightarrow \>
UQFCH)\). In contrapositive the last two implications read: \((\sim
UQFCH \> \Rightarrow \> \> \sim ISC \>)\) and \((\sim INESC \>
\Rightarrow \>\> \sim LC\>\> {\rm or}\> \>\sim SL)\). What prevents
linking these into a single implication is of course the non-entropic
hypothesis, but in any case the two results clearly show what quantum
breakdown entails.

        Suppose that one makes an experimental test of quantum mechanics
and finds a discrepancy between the experiment and the quantum
prediction. This would typically be of the form \(<R>_{exp}^\phi \not =
(\phi,A_R \phi)\) where the left-hand side is the experimental mean
value of some observable \(R\) in the state \(\phi\), and \(A_R\) is the
self-adjoint operator that has been associated to this observable. Now
one may be able to find a self-adjoint operator \(B_R\) such that for
all the states tested one has \(<R>_{exp}^\phi = (\phi,B_R \phi).\) In
this case the discrepancy would most probably be explained by saying
that, for some reason or other, the relation previously established
between the observable \(R\) and \(A_R\) was in fact wrong and the
correct relation should be to the operator \(B_R\). Quantum mechanics
would be preserved, ``discrepancy" would then become ``effect" and
receive some usual explanation within the formalism as due to some new
interaction energy, particle, etc. As such effects accumulate one can
even foresee the possibility of a ``keplerian" revision in which an
equivalent formalism gives an account of the observed physics in a more
elegant simplified form, but this would not be a true subversion of
quantum mechanics as we know it. If however the observed phenomena {\em
cannot \/} be explained by using a different self-adjoint operator,
then, under the assumptions of Section~\ref{sec:theprojection},
superluminal communication becomes reality. Once this happens, causal
paradoxes arise if one is to maintain lorentz covariance. If in addition
the superluminal channel is non-entropic, then one must definitely
abandon either lorentz covariance or the second law. In any case lorentz
covariance is seriously threatened. Barring this, the projection
postulate must be modified. In short, quantum mechanics will either
persist essentially in its present form with absolutely no discernible
deviation, or there will be a radical revision of physics. There cannot
be any small revision. Quantum theory, thermodynamics, and lorentzian
space-time are so interrelated that what affects one affects all.

And independent argument relating the second law of thermodynamics to
locality (absence of superluminal influences) has been brought forward
by Elitzur \cite{elitzur:nohave}, without however expliciting the
non-entropic assumption concerning the mechanisms of the hypothetical
influence.

\section{The covering law and superluminal communication.}
\label{sec:thecovering}

The results of Section~\ref{sec:theprojection} are ``local" in nature in
that they shows that no deviations  of certain types from ordinary
hilbert space quantum mechanics, no matter how small, can be allowed if
the impossibility of superluminal communication is to be maintained. But
one is then immediately led to ask the interesting ``global" question as
to whether the impossibility of superluminal communication can be used
as an axiom leading to hilbert space itself. This would provide a truly
physical basis for hilbert space quantum theory motivated by the
space-time condition of lorentz covariance. In this section we present
some preliminary results in quantum logics that show that such a thesis
is plausible. In a subsequent publication we shall present further
evidence, which does not appeal to signals, connecting lorentz
covariance and the hilbert space structure.  We know from
Piron \cite{piron:foundations} that a generalized hilbert space model can
be constructed if the quantum logic obeys the so-called covering law. It
is this law that we shall try to substantiate by space-time
considerations.  A possible relation between the covering law and the
non existence of superluminal signals was also independently postulated
by Nicolas Gisin (private correspondence) who derived the covering law
from a condition similar to our CNS below.

Now whereas in Section~\ref{sec:theprojection} given the collapse
hypothesis, we derive facts about detectors from the structure of
hilbert space, and the hypothesis of the impossibility of superluminal
communication, to prove anything in the reverse direction seem to be
impossible without positing something about the detection process, for
otherwise we can simply and circularly {\em define\/} detections as
those processes that do not allow long-range correlations to be used to
send superluminal signals. This means that one must append to the axioms
of quantum logic some postulates about the detection process.
Fortunately, enough such expositions already exist in the literature and
we shall simply appeal to some of these introduced by
Guz \cite{guz:filter, guz:anonsymmetric,guz:projection} to make our
point. In this we adopt Guz's notation to facilitate comparison with the
cited works.

We start off conventionally with an orthomodular poset \(L\) of physical
``propositions", and a convex set \(S\) of probability measures on \(L\)
which are to represent ``physical states". The set \(P\) of extreme
points of \(S\) correspond to pure states. We assume that to each pure
state \(p\) there is an atomic ``indicator proposition" \(s(p)\) that
singles it out in that \(p(s(p)) = 1\) and \(q(s(p)) < 1\) for \(q \neq
p\). The map \(p \mapsto s(p)\) is assumed to provides a bijection
between the pure states and the atoms of \(L\).

Measurements are to be described by a ``collapse" scheme pretty much
modeled on the notion of an ideal measurement in the conventional
quantum formalism. In this view, if a proposition \(a\) is tested in a
pure state \(p\) and found to be true, then the state is transformed
(``collapses") into a new pure state \(p_a\). This collapse occurs with
frequency \(p(a)\). This scheme is subject to the following conditions:
\begin{enumerate} \item Propositions that test with certainty do not
collapse the state. That is, \(p_a = p\) whenever \(p(a) = 1\). In
particular, \(p_1 = p\) and \(p_{s(p)}=p\). \item A repeated measurement
yields the same results. That is, \(s(p_a) \leq a\). \end{enumerate} For
two pure states \(p, q\) the number \((p\,:\,q) = p(s(q))\) is called
the {\em transition probability} of \(p\) to \(q\) and corresponds to
the fraction of times \(p\) collapses to \(q\) when tested by the
indicator proposition of this last state. One sees that \((p\,:\,q) =
1\) implies that \(p = q\).

We shall assume that given a set of pair-wise orthogonal propositions
\(\{b_1, \dots, b_n\}\) there is a physical apparatus that tests them
simultaneously with mutually exclusive outcomes. If the state is \(p \in
P\) then the outcome that renders proposition \(b_j\) true and all the
others false occurs with frequency \(p(b_j) = (p\,:\,p_{b_j})\). Note
that in this case the ``detection rate" of the whole apparatus is
\(\sum_{j=1}^n (p\,:\,p_{b_j}) = \sum_{j=1}^n p(b_j) = p(b_1 \vee \cdots
\vee b_n)\). If \(\bigvee_{j=1}^n b_j = 1\), then after the state passes
through the apparatus it becomes the {\em mixed \/} state \[\sum_{j=1}^n
(p\,:\,p_{b_j}) p_{b_j}.\]

Let us now contemplate again an EPR-type space-time situation in which
one has a state \(p\), at one site \(A\) an apparatus corresponding to a
proposition \(a \in L\), and at a distant site \(B\) an apparatus
corresponding to the pair-wise orthogonal set of propositions \(\{b_1,
\dots, b_n\}\) with \(\bigvee_{j=1}^n b_j = 1\). The arrangement is to
operate in such a way that the events corresponding to registries in the
apparatuses are space-like separated. We thus assume that the
propositions \(a\) along with the \(b_j\) form a commuting set. For
there to be no signals from site \(B\) to site \(A\) due to correlations
present in state \(p\) (we will call this the ``no-signal hypothesis"),
the detection rate at \(A\) must be independent of the apparatus used at
site \(B\).

Now it doesn't seem possible to deduce the covering law just from the
no-signal hypothesis. An extension of this hypothesis however to any
situation formally as above where only commutativity (and not just
space-like separation) is assumed does lead to the covering law.
\begin{quote} Commutative no-signal hypothesis (CNS): {\em Let \(a, b_1,
\dots, b_n \in L\) be a commutative set and suppose the \(b_j\)
pair-wise orthogonal with \(\bigvee_{j=1}^n b_j = 1\), then
\(\sum_{j=1}^n (p\,:\,p_{b_j}) p_{b_j}(a)\) is independent of the set
\(\{b_1, \dots, b_n\}\)}. \end{quote} This can be viewed as saying there
is no ``statistical quantum contextualism", that is, the detection rate
of a proposition is independent of what compatible mutually exclusive
and exhaustive set of propositions one has measured just prior to it. If
this set of propositions is measured at a space-like separated site,
this follow from the no signal hypothesis.

Now one particular set of \(b_j\) that one can pick is the singleton
\(\{1\}\) for which the above sum is \((p\,:\,p_1)p_1(a) = p(a)\) and so
the number whose independence is posited has to be \(p(a)\). Let now
\(a\) and \(b\) be two commuting propositions, then from the CNS we
deduce: \[(p\,:\,p_b)p_b(a) + (p\,:\,p_{b'})p_{b'}(a) = p(a)\]
\[(p\,:\,p_{b \wedge a'})p_{b \wedge a'}(a) + (p\,:\,p_{b \wedge a})p_{b
\wedge a}(a) + (p\,:\,p_{b'})p_{b'}(a) = p(a)\] from which using the
fact that \(p_{b \wedge a'}(a) = 0\) and \(p_{b \wedge a}(a) = 1\) we
deduce: \((p\,:\,p_b)p_b(a) = (p\,:\,p_{b \wedge a})\) which is the same
as \[(p : p_b)(p_b\,:\,(p_b)_a) = (p\,:\,p_{a \wedge b})\] which says
that the detection rate with the successive observations of \(b\)
followed by \(a\) is equal to the detection rate of the observation of
\(a \wedge b\). By complete symmetry we also have
\((p\,:\,p_a)(p_a\,:\,(p_a)_b) = (p\,:\,p_{a \wedge b})\).

Consider now \(b = s(q_a)\) where \(q\) is any other state. As \(b \leq
a\), the two commute and we can apply the above results. Now, since
\(b\) is an atom, for any state \(r\) one has \(r_b = s^{-1}(b) = q_a\)
provided \(r(b) \neq 0\). Assume \(p(b) \neq 0\), then by the last
formula of the previous paragraph one deduces \((p\,:\,p_a)
(p_a\,:\,q_a) = (p\,:\,q_a)\). If now \((p\,:\,p_a) = (p\,:\,q_a)\) one
concludes that \((p_a\,:\,q_a) = 1\) which implies that \(p_a = q_a\)
thus one has shown \[(p\,:\,p_a) = (p\,:\,q_a) \neq 0 \Rightarrow p_a =
q_a.\]

What is interesting about this result is that this is precisely the
property one needs in the axiomatic scheme presented by
Guz \cite{guz:projection} to deduce the covering law for the poset \(L\).
Thus if one can somehow justify the extension of necessary conditions on
space-like separated measurement to generally commuting measurements,
the covering law would follow from the no-signal hypothesis. Lacking
this, we have a weaker result in that a stronger hypothesis (CNS), leads
to the covering law in at least one existing axiomatic framework. This
axiomatics was not, to our awareness, conceived to establish a
connection between space-time structure and the covering law and as such
may not be the best to make such a connection plausible. We shall
address this problem in a future publication. There is a general
weakness in all present axiomatic schemes for quantum mechanics, they do
not distinguish space-like separated propositions from other commutative
pairs, for the only relation that one normally posits between space-time
and quantum logic is that space-like separation leads to commutativity.
Only by a joint axiomatization of both lorentzian space-time and hilbert
space quantum mechanics can one hope for anything better. Now
commutativity is generally interpreted as ``commensurability" and this
means the ability to make simultaneous measurements which in turn
suggests that space-like separations always play some role in
commutativity, yet this has never been thoroughly examined. Furthermore
if one believes in general unitary symmetry in hilbert-space, it's not
hard to find examples in which a commutative pair of observables
pertaining to a single localized system (a bound pair of particles say)
is unitarily equivalent to a pair of observables at the opposite arms of
an EPR apparatus. Such a symmetry would thus extend the requirements on
space-like commutativity to commutativity in general. That arguments
pertaining to space-like commutativity can be carried over to general
commutativity has already been pointed out by Home and
Sengupta \cite{homeetal:nohave} in relation to Bell's inequalities. It
may thus well be that the weaker result with the CNS assumption is not
too distant from the desired strong result especially if one imposes
strong symmetry requirements. In any case, it is clear that one cannot
hope to reach a full understanding of hilbert-space quantum mechanics
without linking it to space-time structure.

\section{Conclusions} In  previous sections we gave some arguments in
support of the thesis that, given certain hypotheses, standard quantum
mechanics is structurally unstable, and neighboring theories generically
run into difficulties with relativity. This spells trouble for those
that feel that standard quantum mechanics is only an approximation and
propose alternatives. Great care must be taken in formulating these
alternatives. Some have even come to feel that all alternatives are
ruled out. Stephen Weinberg in his book  \cite{weinberg:dreams} has
tentatively reached this conclusion: \begin{quote} \ldots I could not
find any way to extend the nonlinear version of quantum mechanics to
theories based on Einstein's special theory of relativity \ldots  both
N. Gisin in Geneva and my colleague Joseph Polchinski at the University
of Texas independently pointed out that \ldots the nonlinearities of the
generalized theory {\em could\/} be used to send signals instantaneously
over large distances\ldots\ At least for the present I have given up on
the problem; I simply do not know how to change quantum mechanics by a
small amount without wrecking it altogether.

This theoretical failure to find a plausible alternative to quantum
mechanics, \ldots suggest to me that quantum mechanics is the way it is
because any small change in quantum mechanics would lead to logical
absurdities. If this is true, quantum mechanics may be a permanent part
of physics. Indeed, quantum mechanics may survive not merely as an
approximation to a deeper truth, \ldots but as a precisely valid feature
of the final theory. \end{quote}

A careful analysis of what really goes into the perception of this
rigidity reveals the striking role played by the projection hypothesis
or some modification thereof. This hypothesis in turn is based on the
notion of instantaneous physical state which obviously is a notion tied
to a given inertial frame. It is this frame-dependence that cannot be
reconciled with relativity in the alternative theories. Abandoning such
a frame-dependent notion would mitigate arguments against alternatives
and opens up a true possibility for changing quantum mechanics ``by a
small amount" (as far as numerical predictions are concerned) without
``wrecking it altogether" (maintaining relativity). Weinberg's reasons
for his final speculation on the survival of quantum mechanics may in
the end not be all that compelling.

In recent years there has emerged a new quantum mechanical formalism
(along with several interpretations of it) based on consistent histories
and decoherence  \cite{foobar, goobar}, that essentially does away with
the projection postulate and to a large extent with the notion of
instantaneous state. As such, it escapes the analysis of this paper and
shows a very promising possibility for those that wish to modify quantum
mechanics and still maintain special relativity. We shall investigate
these questions in a future publication.

\section{Acknowledgments}

I wish to thank Nicolas Gisin for helpful correspondences. This research
was supported by the Minist\'erio de Ci\^encia e Tecnologia (MCT) and
the Conselho Nacional de Desenvolvimento Cient\'\i fico e Tecnol\'ogico
(CNPq).

\appendix \section{Appendix: The theorem}\label{sec:thetheorem}

        We start with some notation and preliminaries. Our hilbert
spaces are complex and finite dimensional. For a hilbert space \(H\) we
denote by \(L(H)\) the space of operators in \(H\), and by \(I\) the
identity map. If \(e \in H\) we denote by \(\vert e \vert\) the
equivalence class of \(e\) under the relation by which two elements
\(e\) and \(f\) are equivalent if and only if \(e = \theta f\) for some
unimodular complex number \(\theta\). We denote by \(S_H\) the resulting
quotient space. A positive rank-one operator is necessarily of the form
\((e,\cdot)e\) with \(e\) determined only up to a unimodular multiple.
Thus the space of positive rank-one operators is in a bijective
correspondence with \(S_H\) minus the class of the zero vector. We shall
reserve the Greek letter \(\rho\) to represent a positive rank-one
operator.  A map \(G:S_H \to S_K\) can be lifted in infinitely many ways
to a map \(\hat G : H \to K\) by associating to \(e \in H\) an
arbitrarily chosen element of the equivalence class \(G \vert e \vert\).
We say that \(\hat G \) {\em represents\/} \(G\). Two such
representations \(\hat G_1 \) and \(\hat G_2 \) are related by \(\hat
G_1  f = \theta(f)\hat G_2  f\) for some unimodular function \(\theta\).
If \(\hat G  \) can be chosen to be linear or antilinear, we say that
\(G\) is {\em linearly\/} or {\em antilinearly representable\/}. We
reserve the Greek letter \(\theta\) to indicate unimodular functions.
We denote by \(\lbrack f_1, f_2, \dots, f_q \rbrack\) the linear
subspace generated by the indicated vectors. We use the superscript star
symbol \((\cdot)^\star\) to indicate hermitian conjugate and, when the
bar is inconvenient, the complex conjugate of numbers.

\begin{theorem} Let \(H\) and \(K\) be complex finite-dimensional
hilbert spaces and \(W:L(H) \to L(K)\) a linear map. Suppose W maps a
positive rank-one operators to either a positive rank-one operators or
zero, then its action on positive rank-one operators is either of the
form \(W\rho = C\rho C^\star\) where \(C:H \to K\) is linear or
antilinear, or \(W\rho = Tr(D\rho )(k,\cdot)k\) for some positive
operator \(D \in L(H)\) and \(k \in K\). \end{theorem} We shall call the
three kinds of transforms as being of the {\em linear, antilinear\/} and
{\em degenerate\/} type.

This theorem is a relative of Wigner's theorem \cite{bargmann:note} which
we use in part of the proof.

{\em Proof\/}: If \(k_1,\dots ,k_n\) is an orthonormal basis for \(K\),
then one can write the action of \(W\) as \[WA = \sum  Tr(AL_{ij})(k_i,
\cdot )k_j\] for some \(L_{ij} \in L(H)\).

The action of \(W\) on rank-one positive operators can be considered as
a map \(S_H \to S_K\). Let \(T:H \to K\) be a map,  that represents this
action. One easily shows that replacing \(L_{ij}\) with \({1 \over
2}(L_{ij}+L_{ji}^\star)\) one has the same action on \(S_H\) and so we
can assume that \(L_{ij}^\star = L_{ji}\), in particular the diagonal
elements \(L_{ii}\) are self adjoint. We have \(|(k_i,Tf)|^2 =
(f,L_{ii}f)\), so each \(L_{ii}\) is a positive operator and if we let
\(L = \sum L_{ii}\), we have \(\Vert Tf\Vert ^2 = (f,Lf)\). By the
spectral theorem, there is a positive invertible operator \(M\) such
that \(MLM\) is an orthogonal projector. By working with \(TM\) instead
of \(T\), one can assume that \(L\) is a projector and we do so.

Now if \(\dim H = 0\) there is nothing to investigate. If \(\dim H =
1\), generated by the unit vector \(e\), then the \(L_{ij}\) are plain
numbers. By performing a unitary change of basis in \(K\), the matrix
\((L_{ij})\) undergoes a unitary similarity transformation and since it
is self-adjoint, can be brought into diagonal form. Since its rank must
be at most one, it is either zero and there is nothing more to consider,
or it has a diagonal form with \(L_{11} = 1\) and the rest of the
elements zero. In this last case we have \(W\rho = C\rho C^\star\) where
\(C:H \to K\) is the linear map defined by \(Ce = k_1\) and the theorem
is true in this case.

Thus we next investigate the case that \(\dim H  = 2\). If \(\dim K <
2\) we can extend \(K\) by a direct summand to achieve dimension \(2\),
and extend \(T\) by zero into this summand. The case of \(\dim K > 2\)
we leave for later. For \(\dim K = 2\) we choose orthonormal bases in
\(H\) and \(K\) and identify both spaces with \(\lC^2\).

One now has: \[(L_{ij}) = \pmatrix {\alpha I+\vec\beta\cdot\vec\sigma &
AI+\vec B\cdot\vec\sigma\cr A^\star I+ \vec
B^\star\cdot\vec\sigma&\gamma I +\vec\delta\cdot\vec\sigma\cr}\] where
\(\vec\sigma\) are the usual Pauli spin matrices. Now \(L_{11}\) and
\(L_{22}\) sum to a projector. If this is \(I\) then \(L_{22} = I -
L_{11}\), if of rank one, then,being positive, both \(L_{11}\) and
\(L_{22}\) must be proportional to it, and if zero, both must be zero.
In any case there is a basis of \(H\) in which both are diagonal and so
we can assume that both \(\vec\beta\) and \(\vec\delta\) have only the
third component non zero which we shall at an appropriate point denote
simply by \(\beta\) and \(\delta\). Let \(u\) be a vector in \(\lC^2\)
of norm one, then \(u^\star\vec\sigma u\) is a unit vector in \(\lR^3\)
and any such vector can be so constructed. Denoting such a vector by
\(\hat n\) one has that \((Tu,\cdot)Tu\), in the chosen basis, is given
by the following matrix: \[\pmatrix {\alpha +\vec\beta\cdot \hat n &
A+\vec B\cdot \hat n\cr A^\star + \vec B^\star\cdot \hat n&\gamma
+\vec\delta\cdot \hat n\cr}\] This matrix must have rank at most one for
any choice of \(\hat n\) and so it's determinant must vanish. This gives
\[(\alpha+\vec\beta\cdot \hat n)(\gamma +\vec\delta\cdot\hat n)=( A+\vec
B\cdot \hat n)( A^\star + \vec B^\star\cdot \hat n) \] for all \(\hat
n\). In deriving conclusions from this expression one must be careful to
symmetrize the coefficients of the quadratic terms in \(\hat n\) and to
subtract from them the trace contribution since one has \(\hat n \cdot
\hat n = 1\). This now gives us the following relations:
\begin{eqnarray*} \alpha\gamma + {1\over  3} \vec\beta\cdot\vec\delta
&=& \vert A \vert ^2 + {1\over 3}\vec B^\star\cdot\vec B,\\ \alpha \vec
\delta + \gamma \vec\beta &=& A^\star\vec B + A \vec B^\star,\\ \beta_i
\delta_j + \delta_i \beta_j - {2\over 3} \vec\beta\cdot\vec\delta
\delta_{ij} &=& B_i^\star B_j +B_i B_j^\star - {2\over 3} \vec B^\star
\cdot \vec B \delta_{ij}. \end{eqnarray*} If now \(L\) is zero then both
\(L_{11}\) and \(L_{22}\) are zero and we deduce from the first equation
that \(L_{ij} = 0\) and there is nothing more to prove. We now use the
fact that \(\beta_i = \beta \delta_{i3}\) and \(\delta_i = \delta
\delta_{i3}\) The above equations now give rise to: \begin{eqnarray}
\label{eq:1} \alpha\gamma + {1\over 3} \beta\delta &=&\vert A \vert ^2 +
{1\over 3}\vec B^\star\cdot\vec B,\\ \label{eq:2} \alpha\delta + \gamma
\beta &=& 2 \re (A^\star B_3),\\ \label{eq:3} \re (A^\star B_i) &=&
0,\qquad \qquad \qquad \qquad \qquad (i \not = 3),\\ \label{eq:4} 2
\beta\delta  &=& 2 \vert B_3 \vert^2 -  \vert B_1 \vert^2 - \vert B_2
\vert ^2,\\ \label{eq:5} \re (B_i^\star B_3) &=& 0, \qquad \qquad \qquad
\qquad \qquad (i \not = 3)\\ \nonumber -\beta\delta &=& 2\vert B_1 \vert
^2 - \vert B_2\vert^2 - \vert B_3\vert^2,\\ \label{eq:6} &=& 2\vert B_2
\vert ^2 - \vert B_1\vert^2 - \vert B_3\vert^2,\\ \label{eq:7} \re
(B_1^\star B_2) &=& 0. \end{eqnarray}

From (\ref{eq:6}) one deduces that \(\vert B_1 \vert = \vert B_2
\vert\). Now one still has the freedom to make a unitary change of basis
in  \(H\) by a diagonal matrix with unimodular entries. The effect of
this is to perform a {\em real\/} rotation of the \((B_1,B_2)\) vector.
In this way one can choose the imaginary part of \(B_1\) to be rotated
to zero. Thus we assume \(B_1 = B\) a real number, and by (\ref{eq:7})
one has \(B_2 = \pm iB\). There are now two cases:

{\em Case I:\/} \(B \not = 0\)

From (\ref{eq:3}) we have \(A = 0\) and from (\ref{eq:5}) \(B_3 = 0\).
Equations (\ref{eq:4}) and (\ref{eq:6}) are now equivalent and give
\(\beta\delta = -B^2\) and this in equation (\ref{eq:1}) gives
\(\alpha\gamma = B^2\). Using these in equation (\ref{eq:2}) we deduce
that \(\alpha^2 = \beta^2\) and \(\gamma^2 = \delta^2\). Now since the
\(L_{ii}\) are positive operators we have that \(\alpha \geq 0\) and
\(\gamma \geq 0\). The projector L has the
matrix:\[\pmatrix{\alpha+\gamma+\beta+\delta&0\cr
0&\alpha+\gamma-\beta-\delta\cr}.\] Since \(B \not=0\) neither
\(\alpha\) nor \(\gamma\) is zero. Using now the fact that \(\alpha\)
and \(\gamma\) are positive, that \(\beta\) and \(\delta\) have opposite
signs, and that \(\alpha^2 = \beta^2\) and \(\gamma^2 = \delta^2\) we
deduce that neither diagonal entry in the matrix can be zero. Thus L
must be the identity. This means that \(\beta = - \delta\) and so all
four numbers must have the same modulus. Exchanging the two basis
elements in \(K\) exchanges the role of \(L_{11}\) and \(L_{22}\), thus
without loss of generality we can put \(\alpha = \beta = \gamma =
-\delta = {1 \over 2}\). Changing the sign of one of the basis elements
of \(H\) results in changing the sign of the \((B_1,B_2)\) vector so we
can choose \(B\) to be positive, in which case it must be \({1 \over
2}\) by previous equations. Thus this case reduces to the following
canonical forms: \[{1 \over 2}\pmatrix{I + \sigma_3&\sigma_1 \pm i
\sigma_2\cr \sigma_1 \mp i \sigma_2& I - \sigma_3\cr}\]

{\em Case II:\/} \(B = 0\)

A unitary change of basis in \(K\) causes the representation for
\((L_{ij})\) to change by subjecting the two matrices
\[\pmatrix{\alpha&A\cr A^\star&\gamma\cr }\qquad\pmatrix{\beta&B_3\cr
B_3^\star&\delta\cr}\] to the same unitary similarity transformation.
The two matrices are hermitian, and so there is no loss in generality in
assuming that the second is diagonal and thus \(B_3 =0\). The two now
equivalent equations (\ref{eq:4}) and (\ref{eq:6}) give \(\beta\delta =
\vert B_3 \vert^2= 0\), and then from (\ref{eq:1}) we get \(\alpha\gamma
= \vert A \vert ^2\). There are now two subcases:

{ \em Subcase IIa:\/} \(L\) is rank one.

Without loss of generality assume that the second entry of \(L\) is
zero, then we have \(\alpha + \gamma = \beta + \delta\) and since the
first entry must be one, one has that both sums must be \({1\over 2}\).
Since \(\beta\delta = 0\) one of the two must be zero and the other
\({1\over 2}\). Without loss of generality we set \(\delta = 0\).
Equation (\ref{eq:2}) now gives \(\gamma = 0\) which finally implies
that \(A = 0\) and we have the canonical form: \[{1\over
2}\pmatrix{I+\sigma_3&0\cr 0&0\cr}\]

{\em Subcase IIb:\/} \(L = I\)

We now must have \(\beta + \delta = 0\) which with \(\beta\delta = 0\)
means that both are zero. Thus the second one of the displayed pair of
matrices above is zero and we are free to diagonalize the first one. In
this case, from \(\alpha\gamma = \vert A \vert^2\) one sees that either
\(\alpha\) or \(\gamma\) must be zero and the other 1. Without loss of
generality, assume \(\alpha = 1\), and we have the final canonical form:
\[\pmatrix{I&0\cr 0&0\cr}\]

Let us now examine the cases. One of the forms of Case I corresponds to
\(W\rho = I\rho I\) and the other to \(W\rho  = J\rho J\) where \(J\) is
the antilinear involution on \(\lC^2\) given by \(J(u_1,u_2) = (\bar
u_1,\bar u_2)\). The form of Case IIa corresponds to \(W\rho  =P\rho P\)
where \(P\) is the projector on the first component. The last form
cannot be expressed via a linear or antilinear transformation and
corresponds to \(W\rho  = Tr(\rho )(k_1,\cdot)k_1\). We see here the
three types mentioned at the beginning: the {\em linear, antilinear\/}
and {\em degenerate\/}.

If we take into account the changes of basis that were made, and the
fact that \(L\) was changed into a projector by considering \(TM\)
instead of \(T\) we see that the general (non-canonical) form for the
degenerate type (still with both dimensions two) is \(Tf =
\theta(f)(f,Df)^{1/2}k\) for some positive definite operator \(D\) and
some \(k \in K\).

Consider now the general case of arbitrary dimension greater or equal to
two for each hilbert space.  Choose a two-dimensional orthogonal
projector \(Q\) in \(K\) and a two-dimensional orthogonal projector
\(P\) in \(H\). One has that \(QTP\) is one of the three types. These
local representations will now be joined into a global one.

Firstly we shall keep \(P\) fixed, and by joining the representations of
the \(QTP\), find a representation for \(TP\). For simplicity's sake we
shall generally omit writing \(P\) in the next two paragraphs.

Consider the degenerate type. Suppose that \(K_0 \subseteq K\) is a
subspace with the property that if \(Q_0\) is the orthogonal projector
on \(K_0\), then \(Q_0T f = \theta(f)(f,Df)^{1/2}g\) for some \(g \in
K_0\) and some positive definite operator \(D\). If \(K_0 \not = K\),
choose a vector \(h\) orthogonal to \(K_0\). Let  \(Q\) be the
orthogonal projection onto the subspace \(\lbrack g,h\rbrack\) and
\(Q_g\) the projector onto \(\lbrack g \rbrack\). Suppose that \(QT\) is
either of the linear or antilinear type. Then one has from \(Q_g Q_0Tf =
Q_g QTf\) that \(\theta(f)(f,Df)^{1/2} = \theta'(f)\phi(f)\) for some
linear or antilinear form \(f\). This is obviously impossible due to the
positive definiteness of \(D\). Thus the type must be degenerate and we
have \(QTf = \theta'(f) (f, D' f)^{1/2} k\) for some vector \(k\) and
some positive definite operator \(D'\). By the same token as before we
must now have \(\theta(f)(f,Df)^{1/2}g = \theta'(f) (f, D'
f)^{1/2}Q_gk\). This means that \(g = a Q_gk\) for some non-zero number
\(a\) and \(\theta'(f) (f,D' f)^{1/2} = a \theta(f)(f,Df)^{1/2}\) Let
\(K_1\) be the span of \(K_0\) and \(h\), and \(Q_1\) the orthogonal
projector onto it. From \(Q_1 = Q_0 + Q -Q_g\), we get that \(Q_1Tf =
\theta (f) (f,Df)^{1/2}g + \theta' (f) (f, D'f)^{1/2}(I-Q_g)k =
\theta(f)(f, Df)^{1/2}ak\) which is precisely the form that was assumed
for \(K_0\). By induction therefore, if any of the two-dimensional forms
\(QT\) is degenerate then \(Tf = \theta(f)(f,Df)^{1/2}k\), for some
vector \(k \in K\) and some positive definite operator \(D\).

Suppose now that all the \(QT\) types are linear or antilinear. If the
dimension of the linear span of the range of \(T\) in \(K\) is two or
less then we are through since for some two-dimensional projector \(Q\)
we would have \(TP = QTP\). Suppose therefore that there are at least
three linearly independent vectors  in the range and so choose three
such \(k_1, k_2, k_3\) that are pairwise orthogonal. Let \(Q_{ij}\) be
the orthogonal projection on \(\lbrack k_i,k_j \rbrack\). One has
\(Q_{ij}Tf = \theta_{ij}(f)(\phi_{ij}(f)k_i + \psi_{ij}(f)k_j)\) for
linear or antilinear functionals, as the case may be, \(\phi_{ij}\) and
\(\psi_{ij}\). In particular one has that \(\phi_{12}\) and
\(\psi_{12}\) are linearly independent and so \(\psi_{13}\) is a linear
combination of the two or their complex conjugates, say \(\psi_{13}(f) =
a \phi_{12}(f)+b \psi_{12}(f)\) (the case \(\psi_{13}(f) =
a\overline{\phi_{12}(f)}+b \overline{\psi_{12}(f)}\) is analogous). Now
when \(\phi_{12}(f) = 0\), \(Tf\) is proportional to \(k_2\) and so
\((k_3, Tf) = 0\) which implies that \(\psi_{13}(f) = 0\), and hence \(b
= 0\). One must thus have \(\psi_{13}(f) \) be proportional to
\(\phi_{12}(f)\) or its complex conjugate. A similar reasoning leads to
the conclusion that \(\psi_{23}(f)\) is proportional to \(\psi_{12}(f)\)
or its complex conjugate. But \((k_3,Tf) = \theta_{13}(f)\psi_{13}(f) =
\theta_{23}(f)\psi_{23}(f)\). This proportionality contradicts, by the
linear independence of \(\phi_{12}\) and \(\psi_{12}\), the previously
proved proportionalities unless both \(\psi_{i3}, i= 1,2\) are zero.
This now contradicts the existence of the \(k_i\). Thus the dimension of
the span cannot be greater than two and this case is done.

We now pass to joining the local representations of \(TP\) to a global
one of \(T\).

Suppose that for a two-dimensional projector \(P\) one has that \(TPf =
\theta(f)(f,Df)^{1/2}g\) for some positive definite operator \(D\) and
some vector \(g\). Suppose that the range of \(T\)  is not wholly
contained in a one-dimensional subspace. Then there is a vector \(h\)
with \(Th\) and \(g\) linearly independent. Consider the two dimensional
subspaces \(\lbrack f,h\rbrack\) where \(f \in PH.\) Since on these
subspaces the image of \(T\) contains at least two linearly independent
vectors, the type cannot be degenerate and we must have by either
linearity or antilinearity that \(T(f+\alpha h) = \theta_1 (f,\alpha)Tf
+ \theta_2(f,\alpha)\alpha Th = \theta_1
(f,\alpha)\theta(f)(f,Df)^{1/2}g + \theta_2(f,\alpha)\alpha Th.\) (Note
that a complex number and its complex conjugate differ by a unimodular
factor, so both the linear and antilinear cases are subsumed in the last
expression.) This now defines the action of \(T\) on the three
dimensional subspace \(PH \oplus \lbrack h \rbrack\). The coefficient of
\((g,\cdot)Th\) in \((T(f + \alpha h),\cdot)T(f+\alpha h)\) must, by the
general expression for W, be of the form \((f+\alpha h,N(f+\alpha h))\)
for some linear operator N, and so we must have
\[\overline{\theta_1(f,\alpha)}\overline{ \theta(f)} \theta_2(f,\alpha)
(f,Df)^{1/2}\alpha=\] \[ (f,Nf) + \alpha(f,Nh) + \bar \alpha (h,Nf) +
\vert \alpha \vert ^2 (h,Nh)\] Now the first and fourth term on the
right-hand side must vanish since the left-hand side vanishes when
either \(f\) or \(\alpha\) vanishes. The modulus of the left-hand side
does not change if we multiply \(\alpha\) by a unimodular number, and
for this to be true of the right-hand side then, for a given \(f\), one
of the two middle terms must vanish. Thus as a functional of \(f\), if
one of \((f,Nh)\), \((h,Nf)\) is not the zero functional, then the other
must vanish whenever the first one doesn't, which is only possible for
the zero functional. Thus only one of the middle terms is not zero,
which contradicts the positive definiteness of \(D\) since a linear or
anti-linear functional always has a non-trivial kernel.  Thus in our
case the image of \(T\) is contained wholly in a one-dimensional
subspace \(\lbrack k \rbrack\) and one must have, by the general
expression for \(W\), taking \(k_1 = k\), that \(Tf =
\theta(f)(f,L_{11}f)^{1/2}k\) on all of \(H\).

We are now left in the last case of all the types of \(TP\) being linear
or antilinear. Remember that we can assumed that \(L\) is a projector.
For \(f \in H\), one has  \(f = Lf + (I-L)f\). Assume both terms are not
zero. On the two-dimensional subspace spanned by the two terms,   \(T\)
is either of the linear or antilinear type, thus \(Tf = \theta_1(f)TLf +
\theta_2(f)T(I-L)f\), but \(\Vert T(I-L)f \Vert = \Vert L(I-L)f \Vert =
0\) Thus \(\vert Tf \vert = \vert TLf \vert\) and so \(TL\) represents
the same transformation \(S_H \to S_K\) that \(T\) does. One thus need
only consider \(T\) on \(LH\) and so assume that \(L\) is the identity
(of dimension at least two, as the one-dimensional case was already
settled). With this assumption, one has that on any two-dimensional
subspace, \(T\) is representable by either a linear or antilinear
isometry, and hence, by the polarization identity, preserves the moduli
of inner products. Let \(e_1, \dots, e_n \) be an orthonormal basis for
\(H\), then the \(Te_1,\dots,Te_n\) are orthonormal. By linearity or
antilinearity on two-dimensional subspaces we have \(T(\sum_i^n
\alpha_i e_i ) = T(\alpha_1 e_1 + \sum_2^n  \alpha_i  e_i) =
\theta(\alpha)\alpha_1 T(e_1)+ \theta'(\alpha) T(\sum_2^n  \alpha_i
e_i)\). Continuing in this vein one has \(T(\sum_i^n   \alpha_i e_i ) =
\sum_i^n \theta_i(\alpha) \alpha_i T(e_i)\). Thus the range of \(T\) is
totally contained in the subspace generated by the \(Te_1,\dots,Te_n\).
By working only with the range, we can thus assume \(K\) and \(H\) are
of the same dimension and so we identify them. By Wigner's theorem, any
mapping of a hilbert space into itself which preserves the moduli of
inner products is representable by either a unitary or an antiunitary
operator. This completes the last case and the proof of the theorem.
Q.E.D


\begin{thebibliography}{xx}

\bibitem{einstein:can} Einstein,~A., Podolsky,~B. and Rosen,~N.,  {\em
Physical Review\/}, {\bf 47}, 777 (1935).

\bibitem{bell:ontheeinstein} Bell,~J.~S.,  {\em Physics \/}, {\bf 1},
195 (1964).

\bibitem{redhead:incompleteness} Redhead,~M.,  {\em Incompleteness,
Nonlocality, and Realism: A Prolegomenon to the Philosophy of Quantum
Mechanics\/}, Claredon Press, Oxford (1987).

\bibitem{eberhard:two} Eberhard,~P.~H.,  {\em Nuovo Cimento B\/}, {\bf
38}, 75 (1977).

\bibitem{eberhard:three} Eberhard,~P.~H.,  {\em  Nuovo Cimento B\/},
{\bf 46}, 392 (1978).

\bibitem{ghirardietal:onsome} Ghirardi,~G.~C.  and Weber,~T.,  {\em
Lettere Nuovo Cimento\/}, {\bf 26}, 599 (1979).

\bibitem{ghirardietal:nohave2} Ghirardi,~G.~C.,  Rimini,~A.  and
Weber,~T., {\em Lettere Nuovo Cimento /}, {\bf 27}, 293 (1980).

\bibitem{herbert:nohave} Herbert,~N.,  {\em Foundations of Physics \/},
{\bf 12}, 1171 (1982).

\bibitem{dieks:communication} Dieks,~D.,  {\em Physics Letters A\/},
{\bf 92}, 271 (1982).

\bibitem{milonietal:photons} Milonni,~P.~W. and Hardies,~M.~L.,  {\em
Physics Letters A\/}, {\bf 92}, 321 (1982).

\bibitem{wootersetal:asingle} Wootters,~W.~K. and Zurek,~W.~H.,  {\em
Nature\/}, {\bf 299}, 802 (1982).

\bibitem{aharonovetal:nohave} Aharonov,~Y. and Albert,~D.,  {\em
Physical Review D\/}, {\bf 24}, 359 (1981).

\bibitem{aharonovetal:istheusualii} Aharonov,~Y. and Albert,~D.,  {\em
Physical Review D\/}, {\bf 29}, 228 (1984).

\bibitem{swiftetal:generalized} Swift,~A.~R., and Wright,~R.,  {\em
Journal of Mathematical Physics\/}, {\bf 21}, 77 (1980).


\bibitem{gleason:nohave} Gleason,~A.~M.,  {\em Journal of Mathematics
and Mechanics \/}, {\bf 6}, 885 (1957).

\bibitem{ludwig:nohave} Ludwig,~G.,  {\em Foundations of Quantum
Mechanics\/}, Springer, New York (1983).

\bibitem{piron:foundations} Piron,~C.,  {\em Foundations of Quantum
Physics\/}, W. A. Benjamin, Inc., London (1976).

\bibitem{pearle:reduction} Pearle,~P., {Physical Review D\/}, {\bf 13},
857 (1976)

\bibitem{weinberg:precision} Weinberg,~S., {\em Physical Review
Letters\/}, {\bf 62}, 485 (1989).

\bibitem{weinberg:testing} Weinberg,~S., {\em Annals of Physics (NY)\/},
{\bf 194}, 336 (1989).

\bibitem{gisin:weinberg} Gisin,~N.,  {\em Physics Letters A\/}, {\bf
143}, 1 (1990).

\bibitem{gisin:quantum} Gisin,~N., {\em Physical Review Letters\/}, {\bf
52}, 1657 (1984).

\bibitem{gisin:reply} Gisin,~N., {\em Physical Review Letters\/}, {\bf
53}, 1776 (1984).

\bibitem{gisin:stochastic} Gisin,~N.,  {\em Helvetica Physica Acta\/},
{\bf 62}, 363 (1989).

\bibitem{pearle:comment} Pearle,~P.,  {\em Physical Review Letters\/},
{\bf 53}, 1775 (1984).

\bibitem{pearle:stochastic} Pearle,~P.,  {\em Physical Review D\/}, {\bf
33}, 2240 (1986).

\bibitem{svetlichny:quantum} Svetlichny,~G.,   ``Quantum Evolution and
Space-Time Structure" in H.-D.~Doebner, V.~K.~Dobrev, and P.~Nattermann,
eds., {\em  Nonlinear, Deformed and Irreversible Quantum Systems.
Proceedings of the International Symposium on Mathematical Physics,
Arnold Sommerfeld Institute, 15-19 August 1994, Clausthal, Germany\/},
p. 246, World Scientific, Singapore, 1995, or in a slightly expanded
form on the Los Alamos electronic manuscript archive {\bf
quant-ph/9512004}.

\bibitem{svetlichny:nonlinear} Svetlichny,~G.,  {\em Journal of
Nonlinear Mathematical Physics\/}, {\bf 2}, 2 (1995).

\bibitem{goldin:nohave} Goldin,~G.~A. in {\em Nonlinear, Deformed, and
Irreversible Quantum Systems,\/} ed. by Doebner,~H.-D., Dobrev,~V.~K.
and Nattermann,~P. (World Scientific, Singapore), p. 125 (1995).

\bibitem{doebneretal:nohave1} Doebner,~H.-D. and Goldin,~G.~A.,
``Introducing nonlinear gauge transformations in a family of nonlinear
Schroedinger equations," {\em Physical Review A,\/} (in press).

\bibitem{doebneretal:nohave2} Doebner,~H.-D., and Goldin,~G.~A. and
Nattermann,~P. in {\it Quantization, Coherent States, and Complex
Structures,\/} ed. by Antoine,~J.-P., Ali,~S.~T., Lisiecki,~W.,
Mladenov,~I.~M. and Odzijewicz,~A., (Plenum, New York), p. 27 (1995).

\bibitem{schnorr:nohave} Schnorr,~C.,  {\em Zuf\"alligkeit und
Wahrscheinlichkeit\/}, Lecture Notes in Mathematics, Vol. 218, Springer
Verlag, Berlin (1971).

\bibitem{brillouin:maxwell} Brillouin,~L., {\em Journal of Applied
Physics\/}, {\bf 22}, 334 (1951).

\bibitem{brillouin:physical} Brillouin,~L., {\em Journal of Applied
Physics\/},  {\bf 22}, 338 (1951).

\bibitem{elitzur:nohave} Elitzur,~A.~C., {\em Physics Letters A\/}, {\bf
167}, 335 (1992).

\bibitem{guz:filter} Guz,~W.,  {\em Ann. Inst. Henri Ponincar\'e \/},
{\bf XXIX}, n\({}^o\) 4, 357 (1978).

\bibitem{guz:anonsymmetric} Guz,~W.,  {\em Reports on Mathematical
Physics \/}, {\bf 17}, 385 (1980).

\bibitem{guz:projection} Guz,~W.,  {\em Ann. Inst. Henri Ponincar\'e
\/}, {\bf XXXIV}, n\({}^o\) 4, 373 (1981).

\bibitem{homeetal:nohave} Home,~D. and Sengupta,~S.,  {\em Physics
Letters A\/}, {\bf 102}, 159 (1984).

\bibitem{weinberg:dreams}Weinberg,~S., {\em Dreams of a Final Theory\/},
pp. 88--89, Vintage Books (1992). \bibitem{foobar}Hartle,~J.~B.,
``Spacetime Quantum Mechanics and the Quantum Mechanics of Spacetime",
{\em 1992 Les Houches Ecole d'\'et\'e, Gravitation et Quantifications\/}
\bibitem{goobar}Omn\'es,~R., {\em The Interpretation of Quantum
Mechanics}, Princeton University Press, (1994)

\bibitem{bargmann:note} Bargmann,~V.,  {\em Journal of Mathematical
Physics\/}, {\bf 5}, 862 (1964).


\end{thebibliography}
\end{document}